\documentclass{llncs}
\usepackage{amsmath,amssymb,amsfonts}
\usepackage{mathtools}
\usepackage[usenames,dvipsnames]{color}
\usepackage{xcolor}
\usepackage{xspace}
\usepackage{microtype}
\usepackage{todonotes}
\usepackage{colonequals}
\usepackage{wasysym}
 \usepackage{booktabs}

\usepackage{multirow}
\usepackage{multicol}

\usepackage{paralist}

\usepackage{subfigure}
\usepackage{url}
\usepackage{appendix}
\usepackage{graphicx}
\usepackage{float}
\usepackage{algorithm}
\usepackage{listings}
\usepackage{newalg}
\usepackage{nicefrac}
\usepackage{tikz} 
\usepackage{pgfplots}
\usepackage{wrapfig}

\usetikzlibrary{arrows,decorations.pathmorphing,positioning,fit,trees,shapes,shadows,automata,calc}

\usepackage{macros} % self-defined macros

\title{Safety-Constrained \\ Reinforcement Learning for MDPs\thanks{This work is supported by the Excellence Initiative of the German Research Council and the Sino-German project CAP. UT's work has been partly funded by the awards AFRL \# FA8650-15-C-2546, ONR \# N000141310778, ARO \# W911NF-15-1-0592, and NSF \# 1550212}}

\author{Sebastian Junges\inst{1} \and Nils Jansen\inst{1} \and Christian Dehnert\inst{1}\and\\Ufuk Topcu\inst{2} \and Joost-Pieter Katoen\inst{1}}

\institute{RWTH Aachen University \and University of Texas at Austin}

\begin{document}
\maketitle
%      \ut{Comment by Ufuk}
%      \jpk{Comment by Joost-Pieter}
%      \sj{Comment by Sebastian}
%      \nj{Comment by Nils}
%      \cd{Comment by Chris}

\begin{abstract}
	We consider controller synthesis for stochastic and partially unknown environments in which safety is essential.
Specifically, we abstract the problem as a Markov decision process in which the expected
performance is measured using a cost function that is \emph{unknown} prior to run-time exploration of the state space. 
Standard learning approaches synthesize cost-optimal strategies without guaranteeing safety properties.
To remedy this, we first compute safe, permissive strategies.
Then, exploration is constrained to these strategies and thereby meets the imposed safety requirements.
Exploiting an iterative learning procedure, the resulting policy is safety-constrained and optimal.
We show correctness and completeness of the method and discuss the use of several heuristics to increase its scalability.
Finally, we demonstrate the applicability by means of a prototype implementation. 
\end{abstract}

\section{Introduction}
\paragraph{Probabilistic model checking.}
Many formal system models are inherently stochastic, consider for instance randomized distributed algorithms (where randomization breaks the symmetry between processes), security (e.g., key generation at encryption), systems biology (where species randomly react depending on their concentration), or embedded systems (interacting with unknown and varying environments). These various applications made the \emph{verification} of stochastic systems such as discrete-time Markov chains (MCs) or Markov decision processes (MDPs) an important research topic in the last decade, resulting in several tools like  PRISM~\cite{KNP11}, LiQuoR~\cite{CiesinskiB06}, MRMC~\cite{mrmc} or FMurphi~\cite{PennaIMTZ06}. The always growing set of case studies in the PRISM benchmark suite~\cite{KNP12b} witnesses the applicability of MDP and MC model checking. 

\paragraph{Controller synthesis.}
Contrarily, {controller synthesis} is a relatively new topic in this setting.
Consider a controllable system like, \eg, a robot or some other machine which is embedded into an environment.
Having a formal model of both the controllable entity and the environment, the goal is to synthesize a controller that satisfies certain requirements.
Again, often faithful models are stochastic, imagine, \eg, sensor imprecisions of a robot, message loss, or unpredictable behavior of the environment.
Moreover, it might be the case that certain information---such as cost caused by energy consumption---is not exactly known prior to exploring and observation.

\paragraph{Our problem.}
Given an MDP with a cost structure, synthesize an optimal policy subject to safety constraints. 
This multi-objective model checking problem is studied in~\cite{DBLP:conf/atva/ForejtKP12,DBLP:journals/lmcs/EtessamiKVY08,DBLP:conf/csl/BaierDK14}.
But what if the cost function is not known? 
Consider for instance the following motion planning scenario, placed in a grid-world where a robot wants to move to a certain position.
Acting unpredictably, a janitor moves randomly through the grid.
The robot reaches its goal \emph{safely} if it moves according to a strategy that avoids the janitor.
Moreover, each movement of the robot occasions cost depending on the surface. 
However, the robot only learns the actual costs during physically executing actions within the environment; this requires the exclusion of unsafe behavior prior to exploration.
Consequently, a safe strategy for the robot which simultaneously induces minimal cost is to be found. 

We model robot behavior by an MDP and the stochastic behavior of the environment by a MC.
We are given a \emph{safety condition} specified as a probabilistic reachability objective.
Additionally, we have a \emph{performance condition} bounding the \emph{expected costs} for reaching a certain goal.
A significant problem we are facing is that the costs of certain actions are not known before they are executed.
This calls for using reinforcement learning~\cite{sut98a} algorithms like Q-learning~\cite{qlearning}, where optimal strategies are obtained without prior knowledge about the system.
While this is usually a suitable solution, in this case we have to ensure that no unsafe actions are taken during exploration to ensure an optimal and safe strategy.

\paragraph{Our approach.}
The setting does neither allow for using plain verification nor direct reinforcement learning.
On the one hand, verifying safety and performance properties---in the form of multi-objective model checking---is not possible because the costs of actions are not known.
On the other hand, in practice learning means that the robot will explore parts of the system. 
Doing that, we need to ensure that all unsafe behavior is avoided \emph{beforehand}.
Our solution to these problems is to use the notion of \emph{permissive schedulers}.
In contrast to standard schedulers, where for each system run the next action to take is fixed, more permissiveness is given in the sense that several actions are allowed.
The first step is to compute a \emph{safe} permissive scheduler which allows only safe behavior.
The system is then restricted accordingly and therefore fit for \emph{safe exploration}.

It would be desirable to compute a permissive scheduler which encompasses the set of \emph{all} safe schedulers. 
Having this would ensure that via reinforcement learning a safe scheduler inducing \emph{optimal} cost would obtained.
Unfortunately, there is no efficient representation of such a \emph{maximal permissive scheduler}. 
Therefore, we propose an iterative approach utilizing SMT-solving where a safe permissive scheduler is computed. 
Moreover, the computation can be done via mixed-integer linear programming (MILP).
Out of this, reinforcement learning determines the \emph{locally optimal} scheduler. 
In the next iteration, this scheduler is explicitly excluded and a new permissive scheduler is obtained. 
This is iterated until the performance criterion is satisfied or until the solution is determined to be globally optimal which can be done using known lower bounds on the occurring costs.

\paragraph{Related work.}
In~\cite{draeger-et-al-tacas-2014}, the computation of permissive schedulers  for stochastic 2-player games is proposed for reward properties without additional safety-constraints. 
A dedicated MILP encoding optimizes \wrt to certain \emph{penalties} for actions.
In~\cite{wen-et-al-corr-2015}, permissive safe scheduling is
investigated for transition systems and LTL properties.
Safe or constrained (e.g., by temporal logic specifications) exploration has
also been investigated in the learning literature.
Some recent examples include \cite{abbeel-safe-exploration,jie-pacmdp}.
An overview on safe exploration using reinforcement learning can be found
in~\cite{pecka2014safe}.

\paragraph{Summary of the contributions.}
We give the first approach to controller synthesis for stochastic systems regarding safety and performance in a setting where models are known but cost are not. This encompasses:
\begin{itemize}
	\item an iterative approach on the computation of safe permissive schedulers based on SMT-solving; 
	\item exploitation of permissive schedulers for reinforcement learning towards globally optimal solutions;
%	\item Correctness and completeness of the SMT encoding
%	\item Correctness of the whole iterative method and completeness in the sense that always the optimal solution is found
	\item a discussion of several heuristics to both speed up the computations and avoid too many iterations; and 
	\item a prototype implementation showing promising results on several case studies.
\end{itemize}

\noindent The paper is structured as follows. First, we provide basic notations and formal prerequisites in Section~\ref{sec:preliminaries}. In Section~\ref{sec:permissiveSchedulers} we introduce our notion of permissive schedulers, discuss efficient representations, and introduce technicalities that are needed afterwards. Section~\ref{sec:safeExploration} presents our main results on computing safe and optimal schedulers. After presenting several case studies and benchmark results in Section~\ref{sec:experiments}, we finally draw a conclusion and point to future work in Section~\ref{sec:conclusion}.

%Thoughts:
%
%\begin{itemize}
%	\item Point out different setting in comparison to \cite{draeger-et-al-tacas-2014}
%	\item Compare to other approaches on safe exploration
%	
%\end{itemize}

%
\section{Preliminaries}\label{sec:preliminaries}
In this section, we introduce the required models and specifications considered in this paper, and provide a formal problem statement.

\paragraph{Models.}

For a set $X$, let $2^{X}$ denote the power set of $X$. 
A \emph{probability distribution} over a finite or countably infinite set $\distDom$ is a function $\distFunc\colon\distDom\rightarrow\Ireal$ with $\sum_{\distDomElem\in\distDom}\distFunc(\distDomElem)=\distFunc(\distDom)=1$. 
In this paper, all probabilities are taken from $\Q$.
Let the set of all distributions on $\distDom$ be denoted by $\Distr(\distDom)$.
The set $\supp(\distFunc)=\{x\in\distDom \mid \distFunc(x)>0\}$ is the \emph{support} of $\distFunc \in \Distr(\distDom)$. 
If $\mu(x)=1$ for $x\in\distDom$ and $\mu(y)=0$ for all $y\in\distDom\setminus\{x\}$, $\mu$ is called a \emph{Dirac distribution}.

\begin{definition}{\bf (MDP)}
A \emph{Markov decision process (MDP)} $\MdpInit$ is a tuple with a finite set $S$ of states, a unique initial state $\sinit \in S$, a finite set $\Act$ of actions, and a (partial) probabilistic transition function $\pmdp\colon S\times\Act\rightarrow\Distr(S)$.
\end{definition}
MDPs operate by means of \emph{nondeterministic choices} of actions at each state, whose successors are then determined \emph{probabilistically} \wrt the associated probability distribution.
The set of \emph{enabled} actions at state $s\in S$ is denoted by $\Act(s)=\{a\in\Act\mid\exists\mu\in\Distr(S).\,\mu=\pmdp(s,\alpha)\}$. 
To avoid deadlock states, we assume that $|\Act(s)|\geq 1$ for all $s\in S$.
A \emph{cost function} $\rho\colon S\times\Act\rightarrow\R_{\geq 0}$ for an MDP $\mdp$ adds a cost to each \emph{transition} $(s,a)\in S\times\Act$ with $a\in\Act(s)$.  

A \emph{path} in an $\mdp$ is a finite (or infinite) sequence $\pi=s_0a_0s_1a_1\ldots$ with $\pmdp(s_i, \alpha, s_{i+1}) > 0$ for all $i\geq 0$. 
The set of all paths in $\mdp$ is denoted by $\pathset^\mdp$, all paths starting in state $s\in S$ by $\pathset^\mdp(s)$. 
The cost of finite path $\pi$ is defined as the sum of the costs of all transitions in $\pi$, i.e., $\rho(\pi) = \sum_{i=0}^{n-1} \rho(s_i,a_i)$ where $n$ is the number of transitions in $\pi$.

If $|\Act(s)|=1$ for all $s\in S$, all actions can be disregarded and the MDP $\mdp$ reduces to a \emph{discrete-time Markov chain (MC)}, sometimes denoted by $\dtmc$, yielding a transition probability transition function of the form $\pmdp\colon S\rightarrow\Distr(S)$.
%
%	A probability measure $\pr$ for finite paths of an MC $\dtmc$ is given by the product of transition probabilities:
%	\begin{align*}
%		\pr^\dtmc(s_0,s_1,\ldots,s_n)=\prod_{i=0}^{n-1}\pmdp(s_i)(s_{i+1})
%	\end{align*}
The \emph{unique probability measure} $\pr^\dtmc(\Pi)$ for set $\Pi$ of infinite paths of MC $\dtmc$ can be defined by the usual cylinder set construction, see~\cite{BK08} for details.
The \emph{expected cost} of the set $\Pi$ of paths, denoted by $\expRew^\dtmc(\Pi)$, is defined as $\sum_{\pi \in \Pi} \pr(\pi){\cdot}\rho(\pi)$. 

In order to define a probability measure and expected cost on MDPs, the nondeterministic choices of actions are resolved by so-called \emph{schedulers}\footnote{Also referred to as policies.}. 
As in~\cite{draeger-et-al-tacas-2014}, for practical reasons we restrict ourselves to \emph{memoryless} schedulers; more details about schedulers can be found in~\cite{BK08}.
\begin{definition}{\bf (Scheduler)}\label{def:scheduler}
	A \emph{scheduler} for an MDP $\mdp$ is a function $\sched\colon S\rightarrow\Distr(\Act)$ such that $\sigma(s)(a) > 0$ implies $a \in \Act(s)$. Schedulers using only Dirac distributions are called \emph{deterministic}. The set of all schedulers over $\mdp$ is denoted by $\Sched^\mdp$.
\end{definition}
Deterministic schedulers are functions of the form $\sched\colon S\rightarrow\Act$ with $\sigma(s) \in \Act(s)$. Schedulers that are not deterministic are also called \emph{randomized}. Applying a scheduler to an MDP yields a so-called \emph{induced Markov chain}, as all nondeterminism is resolved. 

\begin{definition}{\bf (Induced MC)}\label{def:induced_dtmc}
	Let MDP $\MdpInit$ and scheduler $\sched\in\Sched^\mdp$. The \emph{MC induced by $\mdp$ and $\sched$} is  $\mdp^\sched=(S,\sinit,\Act,\pmdp^\sched)$ where
	\begin{align*}
		\pmdp^\sched(s,s')=\sum_{a\in\Act(s)} \sched(s)(a)\cdot\pmdp(s,a)(s') \quad \mbox{ for all } s,s'\in S~.
	\end{align*} 
\end{definition}
Intuitively, the transition probabilities in $\mdp^\sched$ are obtained \wrt the random choices of action of the scheduler. 
%When reasProbabilities and expected cost on MDPs are computed based on specific schedulers on the induced MC.
%
\begin{remark}\label{rem:det_schedulers}
\emph{Deterministic schedulers} pick just one action at each state and the associated probability distribution determines the probabilities. In this case we write for all states $s\in S$ and $a\in\Act$ with $\sched(s)(a)=1$:
\begin{align*}
	\pmdp^\sched(s,s')=\pmdp(s,a)(s')\ .
\end{align*}
\end{remark}
%
%This is used to define a probability measure $\pr^{\mdp^\sched}$ \wrt specific schedulers $\sched$ for $\mdp$.

\paragraph{Specifications.}

Specifications are given by combining \emph{reachability properties} and \emph{expected cost properties}. A reachability property $\reachProplT$ with upper probability bound $\lambda\in\Irat$ and target set $T\subseteq S$ constrains the probability to finally reach $T$ from $\sinit$ in $\mdp$ to be at most $\lambda$. 
Analogously, expected cost properties $\expRewProp{\kappa}{G}$ impose an upper bound $\kappa\in\Q$ on the expected cost to reach goal states $G\subseteq S$. 
Combining both typesprovid of properties, the intuition is that a set of bad states $T$ shall only be reached with a certain probability $\lambda$ (safety specification) while the expected cost for reaching a set of goal states $G$ has to be below $\kappa$ (performance specification). 
This can be verified using multi-objective model checking~\cite{DBLP:conf/atva/ForejtKP12,DBLP:journals/lmcs/EtessamiKVY08,DBLP:conf/csl/BaierDK14}, provided all problem data (i.e., probabilities and costs) are a-priori known.

We overload the notation $\finally T$ to denote both a reachability property and the set of all paths that finally reach $T$ from the initial state $\sinit$ of an MC. The probability and the expected cost for reaching $T$ from $\sinit$ are denoted by $\pr(\finally T)$ and $\expRew(\finally T)$, respectively. Hence, $\pr^{\dtmc}(\finally T)\leq\lambda$ and $\expRew^{\dtmc}(\finally G)\leq\kappa$ express that the properties $\reachProplT$ and $\expRewProp{\kappa}{G}$ respectively are satisfied by MC $\dtmc$.

%For a set $T\subseteq S$ of target states of an MC $\dtmc$, we denote the set of finite paths starting in $s\in S$ and ending 
%in the first visit of some $t\in T$ by $\pathset^\dtmc(s,T) = 
%\{s_0 \ldots s_n \in \pathset^{\dtmc}(s) \mid s_n \in T \text{ and } s_i \notin T \text{ for all } i<n\}$.
An MDP $\mdp$ satisfies both reachability property $\varphi$ and expected cost property $\psi$, iff \emph{for all} schedulers $\sched$ it holds that the induced MC $\mdp^\sched$ satisfies the properties $\varphi$ and $\psi$, i.e., $\mdp^\sched\models\varphi$ and $\mdp^\sched\models\psi$.  In our setting, we are rather interested in the so-called \emph{synthesis problem}, where the aim is to find a scheduler $\sched$ such that both properties are satisfied (while this does not necessarily hold for all schedulers).  If $\mdp^\sched\models\varphi$, scheduler $\sigma$ is said to \emph{admit} the property $\varphi$; this is denoted by $\sched\models\varphi$.

%Furthermore, for an MDP $\mdp$, a set of goal state $G$, and a cost function $\rho$, let $\sched^*\in\Sched^\mdp$ denote a scheduler minimizing the expected cost for reaching $G$ from $\sinit$.

%	Notations:
%	\begin{itemize}
%		\item $\sched\models\varphi$, admitting reachability property $\psi=\reachProplT$ for MDP/SG
%		\item maximizing/minimizing scheduler $\sched^*_\psi$ for expected reward property $\psi=\expRewProplT$ in MDP/SG
%		\item $\sched\in\psched$ for scheduler $\sched\in\Sched$ that is ``compliant'' to permissive scheduler $\psched\in\Psched$ (see~\cite{draeger-et-al-tacas-2014})
%	\end{itemize}
%
\iffalse
Finally, there might be cases where schedulers induce loops in an MDP such that target states are not reachable any more. As this is undesired behavior in our setting, we formally fix the so-called \emph{problematic states} where this is possible. For more details, we refer to~\cite{wimmer-et-al-tcs-2014}.
\begin{definition}
	For an MDP $\MdpInit$ and a set of target states $T\subseteq S$, the set of \emph{problematic states} is given by $\probstates = \{s\in\states\mid\exists\sigma\in\Sched.\,\prob[{\mdp^{\sigma}}](\finally T) = 0\}$. The states $S\setminus\probstates$ are called \emph{unproblematic states}.
\end{definition}
\nj{Example for problematic states?}

\sj{Somewhere notice that in our context  With permissive schedulers only these demonic schedulers make sense}
\cd{Can we move this to the encoding section (if it is at all necessary to have it, that is)?}
\fi
%% Let us now state the formal problem we are facing.

\paragraph{Formal problem statement.}
Given an MDP $\mdp_1$ modeling possible controllable behaviors and an MC $\dtmc$ modeling the stochastic behavior of an environment, the synchronous product (see \eg~\cite{pa}) is denoted by $\mdp_1\times\dtmc=\MdpInit$. Let $\rho$ be a cost function over $\mdp$ that is \emph{unknown} to the robot prior to exploring the state space. We assume that for each transition $(s,a)$, the cost is bounded from below and from above, \ie $l_{(s,a)} \leq \rho(s,a) \leq u_{(s,a)}$ with $l_{(s,a)}, u_{(s,a)} \in \Q$ for any $(s,a) \in S \times \Act$.  Let safety specification $\varphi=\reachProplT$ and performance specification $\psi=\expRewProp{\kappa}{G}$ for $\mdp$ with $T,G\subseteq S$.

The \emph{synthesis problem} is to find a scheduler $\sched\in\Sched^\mdp$ such that $\mdp^\sched\models\varphi$ and $\mdp^\sched\models\psi$. The \emph{optimal} synthesis problem is to find a scheduler $\sched^*\in\Sched^\mdp$ such that $\mdp^{\sched^*}\models\varphi$ and $\sched^*$ minimizes the expected cost to reach $G$.

\section{Permissive schedulers}\label{sec:permissiveSchedulers}
As mentioned before, we will utilize the notion of permissive schedulers, where not all nondeterminism is to be resolved.  A permissive scheduler may select a set of actions at each state, such that at a state there might be several possible actions or probability distributions over actions left open. In this sense, permissive schedulers can be seen as sets of schedulers. Here, we discuss properties and efficient representations that are needed later on.
Analogously to schedulers, we consider only memoryless notions.

\begin{definition}{\bf (Permissive scheduler)}\label{def:permissive}
	A \emph{permissive scheduler} of MDP $\MdpInit$ is a function $\psched\colon S\rightarrow 2^{\Distr(\Act)}$. The set of all permissive schedulers for $\mdp$ is $\Psched^\mdp$.
\end{definition}
Intuitively, at each state there is not only one but several distributions over actions available.
\emph{Deterministic} permissive schedulers are functions of the form ${S\rightarrow 2^\Act}$, \ie, there are different choices of action left open.
We use the following notations for connections to (non-permissive) schedulers.
\begin{definition}{\bf (Compliance)}\label{def:permissive_induced}
A scheduler $\sched$ for the MDP $\mdp$ is \emph{compliant} with a permissive scheduler $\psched$, written $\sched\in\psched$, iff for all $s\in S$ it holds that $\sched(s)\in\psched(s)$. 

A permissive scheduler $\psched_{\mathcal{S}}$ for $\mdp$ is \emph{induced} by a set of schedulers $\mathcal{S}\subseteq\Sched^\mdp$, iff for each state $s\in S$ and each distribution $\distFunc\in\psched_{\mathcal{S}}(s)$ there is a scheduler $\sched\in\mathcal{S}$ with $\sched(s)=\distFunc$.
\end{definition}
%

%	\subsection{Safe permissiveness}
We are interested in sets of schedulers that admit our safety specification.

\begin{definition}{\bf (Safe and maximal permissive scheduler)}\label{def:safe_scheduler}
A permissive scheduler $\psched\in\Psched^\mdp$ for the MDP $\mdp$ is \emph{safe} for a reachability property $\varphi=\reachProplT$ if for all $\sched\in\psched$ it holds that $\sched\models\varphi$. The permissive scheduler $\psched$ is called \emph{maximal}, if there exists no scheduler $\sched\in\Sched^\mdp$ with $\sched\not\in\psched$ and $\sched\models\varphi$.
\end{definition}

A safe permissive scheduler contains \emph{only} schedulers that admit the safety specification while a maximal safe permissive scheduler contains \emph{all} such schedulers (and probably more). 
%Note again that we only consider memoryless schedulers here, \ie there might be safe (but not memoryless) schedulers that are not included in the maximal permissive scheduler. 
%\subsection{Representation of permissive schedulers}
Note that even for a set of safe schedulers, the induced permissive scheduler might be unsafe; contradicting choices might evolve, \ie, choosing a certain action (or distribution) at one state might rule out certain memoryless choices at other states; this is illustrated by the following example.
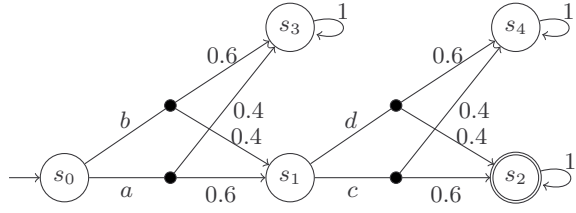
\begin{figure}[t]
	%SHIFT THIS to some other place - dont want to do that while others are working on it..
%\tikzset{>={Latex[width=1.5mm,length=1.5mm]}} 
\centering
\begin{tikzpicture}[scale=1, nodestyle/.style={draw,circle},baseline=(s0)]

    \node [nodestyle, initial, initial text={}] (s0) at (0,0) {$s_0$};
    \node [nodestyle] (s1) [on grid, right=3cm of s0] {$s_1$};
    \node [nodestyle, accepting] (s2) [on grid, right=3cm of s1] {$s_2$};
    \node [nodestyle] (s3) [on grid, above=2cm of s1] {$s_3$};
    \node [nodestyle] (s4) [on grid, above=2cm of s2] {$s_4$};
    
    \node [circle, draw, scale=0.5, fill=black] (s0a) [right=1cm of s0] {};
    \node [circle, draw, scale=0.5, fill=black] (s0b) [above=0.8cm of s0a] {};
    \node [circle, draw, scale=0.5, fill=black] (s1a) [right=1cm of s1] {};
    \node [circle, draw, scale=0.5, fill=black] (s1b) [above=0.8cm of s1a] {};

    \draw (s0) -- node[below] {$a$} (s0a);
    \draw (s0) -- node[above] {$b$} (s0b);
    \draw (s1) -- node[below] {$c$} (s1a);
    \draw (s1) -- node[above] {$d$} (s1b);
    
    \draw[->] (s0a) -- node [below] {$0.6$} (s1);
    \draw[->] (s0a) -- node [right] {$0.4$} (s3);
    \draw[->] (s0b) -- node [right] {$0.4$} (s1);
    \draw[->] (s0b) -- node [above] {$0.6$} (s3);
    
    \draw[->] (s1a) -- node [below] {$0.6$} (s2);
    \draw[->] (s1a) -- node [right] {$0.4$} (s4);
    \draw[->] (s1b) -- node [right] {$0.4$} (s2);
    \draw[->] (s1b) -- node [above] {$0.6$} (s4);
    
    \draw(s4) edge[loop right] node [above] {$1$} (s4);
    \draw(s3) edge[loop right] node [above] {$1$} (s3);
    \draw(s2) edge[loop right] node [above] {$1$} (s2);    
    
   \end{tikzpicture}
	\caption{Example MDP $\mdp$ illustrating conflicting schedulers}
	\label{fig:conflict_schedulers}
\end{figure}

\begin{example}\label{ex_conflicting_choices}
Consider the MDP $\mdp$ depicted in Figure~\ref{fig:conflict_schedulers}, where the only nondeterministic choices occur at states $s_0$ and $s_1$. 
Assume a reachability property $\varphi = \p_{\leq 0.3}(\finally \{ s_2 \})$. 
This property is violated by the deterministic scheduler $\sched_1\coloneqq\{s_0 \mapsto a, s_1 \mapsto c\}$ as $s_2$ is reached with probability 0.36 exceeding the threshold 0.3. 
This is the only unsafe scheduler; removing either action $a$ or $c$ from $\mdp$ leads to a \emph{safe} MDP, \ie the possible deterministic schedulers $\sched_2\coloneqq\{s_0 \mapsto a, s_1 \mapsto d\}$, $\sched_3\coloneqq\{s_0 \mapsto b, s_1 \mapsto c\}$, and $\sched_4\coloneqq\{s_0 \mapsto b, s_1 \mapsto d\}$ are all safe. However, consider the induced permissive scheduler $\psched_{\sched_2,\sched_3,\sched_4}\in\Psched^\mdp$ with $\psched\coloneqq \bigl\{s_0\mapsto \{a,b\},s_1\mapsto\{b,c\}\bigr\}$, where all nondeterministic choices are left open. Unfortunately, it holds that the unsafe scheduler $\sched_1$ is compliant with $\psched_{\sched_2,\sched_3,\sched_4}$, therefore $\psched$ is unsafe.
\end{example}
Example~\ref{ex_conflicting_choices} shows that in order to form a safe permissive scheduler it is not sufficient to just consider the set of safe schedulers. Actually, one needs to keep track that the very same safe scheduler is used in every state. Theoretically, this can be achieved by adding finite memory to the scheduler in order to avoid conflicting actions. 
%Assuming a set $\mathcal{S}$ of safe schedulers, one needs to require for the permissive scheduler $\psched_{\mathcal{S}}$ induced by $\mathcal{S}$ that every scheduler compliant with $\psched_{\mathcal{S}}$ is contained in $\mathcal{S}$.

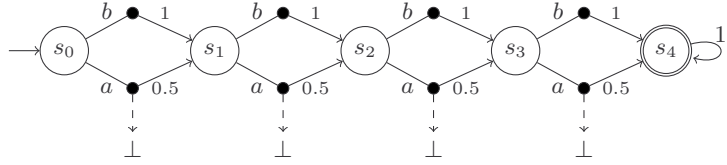
\begin{figure}[h]
	%SHIFT THIS to some other place - dont want to do that while others are working on it..
%\tikzset{>={Latex[width=1.5mm,length=1.5mm]}} 
\centering
\begin{tikzpicture}[scale=1, nodestyle/.style={draw,circle},baseline=(s0)]

    \node [nodestyle, initial, initial text={}] (s0) at (0,0) {$s_0$};
    \node [nodestyle] (s1) [on grid, right=2cm of s0] {$s_1$};
    \node [nodestyle] (s2) [on grid, right=2cm of s1] {$s_2$};
    \node [nodestyle] (s3) [on grid, right=2cm of s2] {$s_3$};
    \node [nodestyle, accepting] (s4) [on grid, right=2cm of s3] {$s_4$};
    
   \node [circle, draw, scale=0.5, fill=black] (s0a) [right=0.5cm of s0, yshift=-1cm] {};
   \node [circle, draw, scale=0.5, fill=black] (s0b) [right=0.5cm of s0, yshift=1cm] {};
   \node [circle, draw, scale=0.5, fill=black] (s1a) [right=0.5cm of s1, yshift=-1cm] {};
   \node [circle, draw, scale=0.5, fill=black] (s1b) [right=0.5cm of s1, yshift=1cm] {};
   \node [circle, draw, scale=0.5, fill=black] (s2a) [right=0.5cm of s2, yshift=-1cm] {};
   \node [circle, draw, scale=0.5, fill=black] (s2b) [right=0.5cm of s2, yshift=1cm] {};
   \node [circle, draw, scale=0.5, fill=black] (s3a) [right=0.5cm of s3, yshift=-1cm] {};
   \node [circle, draw, scale=0.5, fill=black] (s3b) [right=0.5cm of s3, yshift=1cm] {};

	\node [] (s0adummy) [below=0.5cm of s0a] {$\bot$};
    \draw [->,dashed] (s0a) -- (s0adummy);
    
%   	\node [] (s0bdummy) [above=0.5cm of s0b] {$\ldots$};
%    \draw [->,dashed] (s0b) -- (s0bdummy);

	\node [] (s1adummy) [below=0.5cm of s1a] {$\bot$};
    \draw [->,dashed] (s1a) -- (s1adummy);
    
%   	\node [] (s1bdummy) [above=0.5cm of s1b] {$\ldots$};
%    \draw [->,dashed] (s1b) -- (s1bdummy);
    
    \node [] (s2adummy) [below=0.5cm of s2a] {$\bot$};
    \draw [->,dashed] (s2a) -- (s2adummy);
    
%   	\node [] (s2bdummy) [above=0.5cm of s2b] {$\ldots$};
%    \draw [->,dashed] (s2b) -- (s2bdummy);
    
    \node [] (s3adummy) [below=0.5cm of s3a] {$\bot$};
    \draw [->,dashed] (s3a) -- (s3adummy);
    
%   	\node [] (s3bdummy) [above=0.5cm of s3b] {$\ldots$};
%    \draw [->,dashed] (s3b) -- (s3bdummy);

%
%    
    \draw (s0) -- node[below] {$a$} (s0a);
    \draw (s0) -- node[above] {$b$} (s0b);
    \draw[->] (s0a) -- node [below] {\scriptsize$0.5$} (s1);
    \draw[->] (s0b) -- node [above] {\scriptsize$1$} (s1);
    
    \draw (s1) -- node[below] {$a$} (s1a);
    \draw (s1) -- node[above] {$b$} (s1b);
    \draw[->] (s1a) -- node [below] {\scriptsize$0.5$} (s2);
    \draw[->] (s1b) -- node [above] {\scriptsize$1$} (s2);

    \draw (s2) -- node[below] {$a$} (s2a);
    \draw (s2) -- node[above] {$b$} (s2b);
    \draw[->] (s2a) -- node [below] {\scriptsize$0.5$} (s3);
    \draw[->] (s2b) -- node [above] {\scriptsize$1$} (s3);

    \draw (s3) -- node[below] {$a$} (s3a);
    \draw (s3) -- node[above] {$b$} (s3b);
    \draw[->] (s3a) -- node [below] {\scriptsize$0.5$} (s4);
    \draw[->] (s3b) -- node [above] {\scriptsize$1$} (s4);

	\draw(s4) edge[loop right] node [above] {$1$} (s4);

   \end{tikzpicture}
	\caption{MDP $\mdp_{4}$ inducing exponentially many (minimal) conflict sets}
	\label{fig:exponential_size}
\end{figure}

A succinct representation of the maximal permissive scheduler can be gained by enumerating all \emph{minimal} sets of conflicting action choices (now only considering deterministic schedulers), and excluding them from all possible schedulers. We investigate the worst case size of such a set. Assume without loss of generality that for all $s\in S$ the sets $\Act(s)$ are pairwise disjoint.
\begin{definition}{\bf (Conflict set)}
$C \subseteq \Act$ is a \emph{conflict set} for MDP $\mdp$ and property $\varphi$ iff there exists a scheduler $\sched\in\Sched^{\mdp}$ such that $(\forall a\in C.\,\exists s\in S.\,\sched(s)=a)$ and $\sigma\not\models\varphi$. 
 The set of \emph{all conflict sets} for $\mdp$ and $\varphi$ is denoted by $\Confl^\mdp_\varphi$.
 $C\in\Confl^\mdp_\varphi$ is a \emph{minimal} conflict set iff $\forall C'\subsetneq C.\,C'\not\in\Confl^\mdp_\varphi$.
\end{definition}
\begin{lemma}\label{theo:minimalconflict}
The size of the set of all minimal conflict sets for $\mdp$ and $\varphi$ potentially grows exponentially in the number of states of $\mdp$.
\end{lemma}
\paragraph{Proof sketch.}
Let $\mdp_{n} = (S, \sinit, \Act, \pmdp)$ be given by $S = \{ s_0, \ldots, s_{n}, \bot \}$, $\sinit = s_0$, $Act = \{ a_{0}, \ldots, a_{n-1}, b_{0}, \ldots, b_{n-1}, c, d \}$ and
\begin{align*}
	\pmdp(s, \alpha)(t) = \begin{cases}
		0.5 & \text{if } i < n, \alpha = a_{i}, s = s_{i}, t = s_{i+1} \\
		0.5 & \text{if } i < n, \alpha = a_{i}, s = s_{i}, t = \bot \\
		1 & \text{if } i < n, \alpha = b_{i}, s = s_{i}, t = s_{i+1} \\
		1 & \text{if } \alpha = c, s = s_{n}, t = s_{n} \\
		1 & \text{if } \alpha = d, s = \bot, t = \bot \\
		0 & \text{otherwise}
	\end{cases}
\end{align*}
Figure~\ref{fig:exponential_size} shows the instance $\mdp_{4}$ where several copies of the $\bot$-states have been drawn for ease of presentation. 
Consider the property $\varphi = \reachProp{\lambda}{\{s_n^{•}\}}$ with $\lambda = 0.5^{\frac{n}{2} + 1}$. Choosing any combination of $\frac{n}{2}$ of the $b_{i}$ actions yields a minimal conflict set. Hence, there are at least
\begin{align*}
	\binom{n}{\frac{n}{2}} \overset{n:=2m}{=} \frac{(2m)!}{2m!} = \underbrace{\frac{(m+1)}{1} \cdots \frac{2m}{m}}_{m \text{ factors } \geq \; 2} \geq 2^m \overset{m:=\frac{n}{2}}{=} 2^{\frac{n}{2}} \in \Omega\left(\left(\sqrt{2}\right)^{n}\right)
\end{align*}
minimal conflict sets. \hfill $\Box$ \\	

This strongly indicates that an exact representation of the  maximal permissive scheduler is not feasible.
For algorithmic purposes, we strive for a more compact representation. 
It seems natural to investigate the possibilities of using MDPs as representation of permissive schedulers.
Therefore, analogously to induced MCs for schedulers (cf.\ Definition~\ref{def:induced_dtmc}), we define induced MDPs for permissive schedulers.
For a permissive scheduler $\psched\in\Psched^\mdp$, we will uniquely identify the nondeterministic choices of probability distributions $\distFunc\in\psched(s)$ at each state $s\in S$ of the MDP by new actions $a_{s,\distFunc}$.
\begin{definition}{\bf (Induced MDP)} \label{def:induced_mdp}
For an MDP $\MdpInit$ and permissive scheduler $\psched$ for $\mdp$, the \emph{MDP induced by $\mdp$ and $\psched$} is the MDP $\mdp^\psched=(S,\sinit,\Act^\psched,\pmdp^\psched)$ with $\Act^\psched=\{a_{s,\distFunc}\mid s\in S,\distFunc\in\psched(s)\}$ and:
\begin{align*}
	\pmdp^\psched(s,a_{s,\distFunc})(s')=\sum_{a\in\Act(s)}\distFunc(s)(a)\cdot\pmdp	(s,a)(s')\quad
	 \mbox{for $s,s'\in S$ and $a_{s,\distFunc}\in\Act^\psched$}~.
\end{align*}
\end{definition}
Intuitively, we nondeterministically choose between the distributions over actions induced by the permissive scheduler $\psched$. 
Note that if the permissive scheduler contains only one distribution for each state, \ie, in fact the permissive scheduler is just a scheduler, the actions can be discarded which yields an induced MC as in Definition~\ref{def:induced_dtmc}, making this definition backward compatible.

\begin{remark}\label{rem:random_induced}
Each deterministic scheduler $\sched\in\Sched^{\mdp^\psched}$ for the induced MDP $\mdp^\psched$ \emph{induces a (randomized) scheduler} for the original MDP $\mdp$.
In particular, $\sched$ induces a scheduler $\sched'\in\psched$ for $\mdp$ which is compliant with the permissive scheduler $\psched$: For all $s\in S$ there exists an action $a_{s,\distFunc}\in\Act^\psched$ such that $\sched(s)=a_{s,\distFunc}$. The randomized scheduler $\sched'$ is then given by $\sched'(s)=\mu$ and it holds that
\begin{align*}
	\sum_{a \in\Act(s)}\sched'(s)(a)\cdot\pmdp(s,a)(s')=\pmdp^\psched(s,a_{s,\distFunc})(s')\ .
\end{align*}
\end{remark}
	
\begin{remark}\label{rem:submdp}
	A \emph{deterministic permissive scheduler} $\psched_{\text{det}}\in\Psched^\mdp$ for the MDP $\mdp$ simply \emph{restricts} the nondeterministic choices of the	original MDP to the ones that are chosen with probability one by $\psched_{\text{det}}$. The transition probability function $\pmdp^{\psched_{\text{det}}}$ of the induced MDP $\mdp^{\psched_{\text{det}}}$ can be written as
\begin{align*}
	\pmdp^\psched(s,a_{s,\distFunc})(s')=\pmdp	(s,a)(s')\quad
	\mbox{for all $s\in S$ and $a_{s,\distFunc}\in\Act^{\psched_{\text{det}}}$ with $\distFunc(a)=1$}~.
\end{align*}	
	
The induced MDP $\mdp^\psched$ can be seen as a sub-MDP $\mdp^{\mathit{sub}}=(S,\sinit,\Act,\pmdp^{\mathit{sub}})$ of $\mdp$ by omitting all actions that are not chosen. Hence, for all $s,s'\in S$:
\begin{align*}
	\pmdp^{\mathit{sub}}(s,a)(s')=
	\begin{cases}
		\pmdp(s,a)(s') & \text{ if }\exists\distFunc\in\psched(s).\,\mu(a)=1 \\
						0 & \text{ otherwise .}
	\end{cases}
\end{align*}
\end{remark}

%Note that using this representation of permissive schedulers, again for a safe permissive scheduler there might conflicting choices in the induced MDP. This is illustrated by our running example.

\begin{example}\label{ex:submdp}
Recall Example~\ref{ex_conflicting_choices}. The MDP $\mdp^{\psched}$ induced by the \emph{safe} permissive scheduler $\psched$ is the same as $\mdp$, as all available choices of actions are included (see Example~\ref{ex_conflicting_choices}). Note that we use the simplified notation from Remark~\ref{rem:submdp}.
However, consider the safe (but not maximal) permissive scheduler $\psched_{\textit{safe}}$ formed by $\{s_0 \mapsto a, s_1 \mapsto d\}$ and $\{s_0 \mapsto b, s_1 \mapsto d\}$. The induced MDP is the sub-MDP $\mdp^{\psched_{\textit{safe}}}$ of $\mdp$ depicted in Figure~\ref{fig:safe_mdp}.  This sub-MDP has no scheduler $\sigma$ with $\sigma \not\models \varphi$.
\end{example}

\begin{figure}[t]
	%SHIFT THIS to some other place - dont want to do that while others are working on it..
%\tikzset{>={Latex[width=1.5mm,length=1.5mm]}} 
\centering
\begin{tikzpicture}[scale=1, nodestyle/.style={draw,circle},baseline=(s0)]

    \node [nodestyle, initial, initial text={}] (s0) at (0,0) {$s_0$};
    \node [nodestyle] (s1) [on grid, right=3cm of s0] {$s_1$};
    \node [nodestyle, accepting] (s2) [on grid, right=3cm of s1] {$s_2$};
    \node [nodestyle] (s3) [on grid, above=2cm of s1] {$s_3$};
    \node [nodestyle] (s4) [on grid, above=2cm of s2] {$s_4$};
    
    \node [circle, draw, scale=0.5, fill=black] (s0a) [right=1cm of s0] {};
    \node [circle, draw, scale=0.5, fill=black] (s0b) [above=0.8cm of s0a] {};
%    \node [circle, draw, scale=0.5, fill=black] (s1a) [right=1cm of s1] {};
    \node [circle, draw, scale=0.5, fill=black] (s1b) [right=1cm of s1,yshift=1.6cm] {};

    \draw (s0) -- node[below] {$a$} (s0a);
    \draw (s0) -- node[above] {$b$} (s0b);
%	\draw (s1) -- node[below] {$c$} (s1a);
    \draw (s1) -- node[above] {$d$} (s1b);
    
    \draw[->] (s0a) -- node [below] {$0.6$} (s1);
    \draw[->] (s0a) -- node [right] {$0.4$} (s3);
    \draw[->] (s0b) -- node [right] {$0.4$} (s1);
    \draw[->] (s0b) -- node [above] {$0.6$} (s3);
    
%    \draw[->] (s1a) -- node [below] {$0.6$} (s2);
%    \draw[->] (s1a) -- node [right] {$0.4$} (s4);
    \draw[->] (s1b) -- node [auto] {$0.4$} (s2);
    \draw[->] (s1b) -- node [auto] {$0.6$} (s4);
    
    \draw(s4) edge[loop right] node [above] {$1$} (s4);
    \draw(s3) edge[loop right] node [above] {$1$} (s3);
    \draw(s2) edge[loop right] node [above] {$1$} (s2);    
    
   \end{tikzpicture}
	\caption{Induced MDP $\mdp_{\psched_{\textit{safe}}}$}
	\label{fig:safe_mdp}
\end{figure}

\section{Safety-constrained reinforcement learning}
\label{sec:safeExploration}

Recall that the synthesis problem amounts to determining a scheduler $\sigma^*$ of the MDP $\mdp$ such that $\sigma^*$ admits the safety specification $\varphi$ and minimizes the expected cost (of reaching $G$).  A naive approach to this problem is to iterate over all safe schedulers $\sigma_1, \sigma_2, \sigma_3, \ldots$ of $\mdp$ and pursue in the $j$-th iteration as follows.  Deploy the (safe) scheduler $\sigma_{j}$ on the robot.  By letting the robot safely explore the environment (according to $\sigma_{j}$), one obtains the expected costs $c_{j}$, say, of reaching $G$ (under $\sigma_{j}$).  By doing so for all safe schedulers, one obtains the minimum cost.  After checking all safe schedulers, we have obtained a safe minimal one whenever some $c_n$ is below the threshold $\kappa$.  The solution to the synthesis problem is then the scheduler $\sigma_n$ for which $c_n$ is minimal.  Otherwise, we can conclude that the synthesis problem has no solution.  Note that while deploying the safe schedulers, the robot explores more and more possible trajectories, thus becoming more knowledgeable about the (a-priori) unknown cost structure of the MDP.

%We start by considering the following naive approach to our synthesis problem: Iterating over all schedulers, we check whether the scheduler fulfills the safety constraints, and if so, we deploy the scheduler on our robot. We can then \emph{safely} let the robot explore the environment and learn the a-priori unknown performance. Eventually, we either obtain a scheduler which admits our performance criterion or we can conclude that there is no safe scheduler left to consider. 

Although this approach is evidently sound and complete, the number of deployments is excessive.   Our approach avoids this blow-up by:
\begin{enumerate}
\item 
Testing permissive (\ie \emph{sets} of) schedulers rather than one scheduler at a time.  This is done by employing reinforcement learning.
\item
Using that the expected costs $c^*$ under $\sigma^*$ cannot be smaller than the minimal expected cost $c$ in the MDP $\mdp$ (possibly achieved by some unsafe scheduler).  This allows for deciding minimality of scheduler $\sigma_{j}$ by checking $c_j = c$, possibly avoiding exploration of any further schedulers. 
\item
Preventing the deployment of safe scheduler $\sigma_{j}$ whenever the minimal expected cost $c_i$ of all schedulers checked so far ($i < j$) is smaller than the expected cost under $\sigma_{j}$.
\end{enumerate}

%In this section, we improve on this naive iterative approach by reducing the number of deployments required. We start with three observations.
%\medskip

%\noindent\emph{Learning can handle permissive schedulers.}
%We use induced MDPs as representation for permissive schedulers, see Definition~\ref{def:induced_mdp}. This enables to deploy multiple \emph{safe} schedulers at a time. Reinforcement learning is then directly applied to these MDPs while safety is ensured.
%\smallskip
%
%\noindent\emph{Lower bound on the safe optimum.}
%Observe that the performance of the best (potentially unsafe) scheduler is a lower bound for the performance of the best safe scheduler. Moreover, we are initially given lower (and upper) bounds on cost of actions.
%Therefore, we can use model checking to determine the expected cost to reach $G$ on the \emph{original MDP} and abort the iterative approach as soon as we have found a safe scheduler achieving the same performance. While learning more information, we can refine the lower bound after each iteration.
%\smallskip
%
%\noindent\emph{Upper bound on the safe optimum.}
%The best scheduler found so far induces an upper bound on the performance as it is optimal for the already learned parts of the MDP. After computing a new candidate (permissive) scheduler, we can re-compute its performance using the lower bounds on actions on the original MDP. If it does not (potentially) admit a better performance, it does not need to be deployed at all.
%\smallskip

%
\begin{figure}[t]
	\centering
\begin{tikzpicture}
\tikzstyle{outer}= [draw, text centered, shape=rectangle, text width=8cm]
\tikzstyle{inner}=[draw, text centered, shape=rectangle, rounded corners, text width=4cm, minimum height=1.1cm, inner sep=5pt]			

%inputs
\node[outer] (input) {MDP $\mdp$, minimally initialized cost function $\rho$,\\ safety specification $\varphi$, performance specification $\psi$
};
			
%interiour

\node[inner, below=0.7cm of input,xshift=-3.3cm] (computeScheduler) {1. \emph{Compute safe permissive scheduler} $\psched\in\Psched^\mdp$; exclude all previously computed schedulers};

\node[inner, below=0.9cm of computeScheduler, on grid] (learning) {2. \emph{Obtain locally cost-optimal scheduler} $\sched\in\psched$ and refine cost function $\rho$ via reinforcement learning};

\node[inner, below=0.9cm of learning, on grid] (check) {4. \emph{Check} if $\sigma\models\psi$ or if $\sigma$ is optimal};

\node[inner, right=2cm of check, on grid, yshift=1cm] (lowerBound) {3. Compute scheduler $\sched^l\in\Sched^{\mdp}$ on the original MDP $\mdp$ inducing a \emph{lower bound on the expected cost}};

%outputs

\node[outer, below=0.8cm of check, on grid, text width=2cm] (result) {Return $\sigma$};

%big box
\node [draw, dashed, fit = (computeScheduler) (learning) (learning) (check) (lowerBound), inner sep = .2cm] {};

%\node [draw=black, fit = (input) (computeScheduler) (learning) (learning) (check) (lowerBound) (result), inner sep = .2cm] {};

%arrows
\draw (input) edge[-latex', thick] (computeScheduler.north);

\draw (computeScheduler) edge[-latex', thick] node[auto] {Induced MDP $\mdp^
\psched$} (learning);

\draw (learning) edge[-latex', thick] node[auto] {Scheduler $\sigma$} (check);

\draw (learning.east) edge[-latex', thick] node[near end, above=.4cm] {Cost function $\rho$} (lowerBound);

\draw (lowerBound) edge[-latex', thick] node[auto,below=.2cm] {Scheduler $\sched^l$} (check);

\draw (check) edge[-latex', thick] node[pos=0.7,auto] {yes} (result);

\draw (check.west) edge[-latex', thick, bend left] node[auto] {no} (computeScheduler.west);
					
\end{tikzpicture}
	\caption{Overview of safety-constrained reinforcement learning}
	\label{fig:overview_synthesis}
\end{figure}
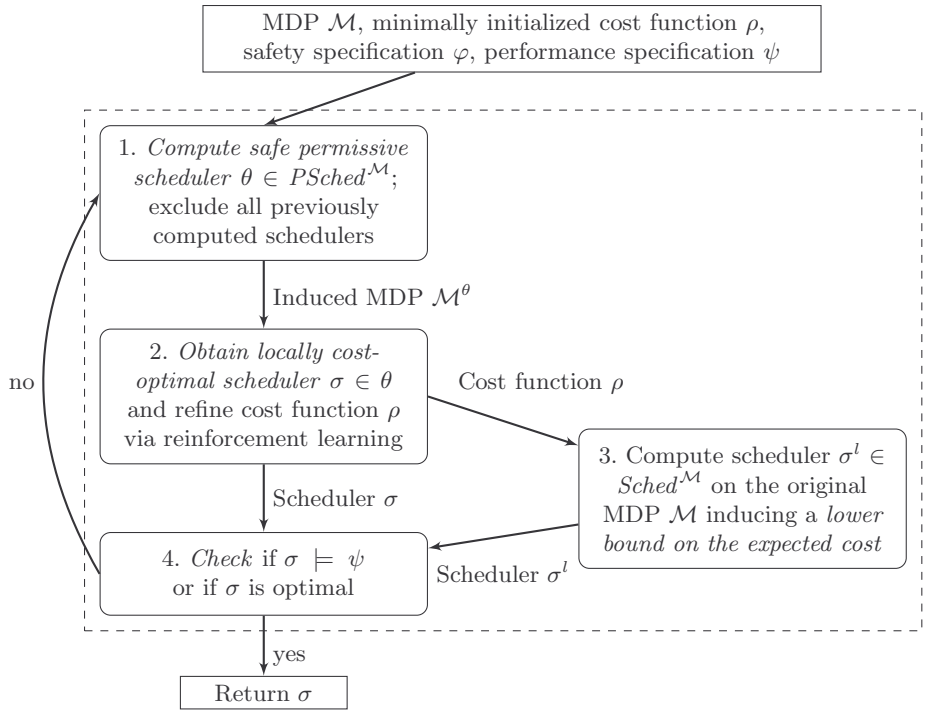

Let us now briefly explain our approach to synthesize a safe and optimal scheduler; further details are given in the rest of this section.  Figure~\ref{fig:overview_synthesis} surveys the approach. We initialize the cost function of the MDP by setting the cost of transition $(s,a)$ to its lower bound $l_{(s,a)}$.  The iterative synthesis of a safe and optimal scheduler is done by iteratively considering permissive schedulers $\theta_1, \theta_2, \theta_3, \ldots$ according to which the MDP $\mdp$ is explored.  This yields a scheduler $\sigma$ whose expected cost is minimal among the schedulers deployed so far.  This search is finished whenever either the expected costs under $\sigma$ is below $\kappa$, $\sigma$ is globally optimal, or no further permissive schedulers can be found.  In the $j$-th iteration, the following four steps are carried out:
\begin{enumerate}
\item 
Determine the $j$-th safe permissive scheduler $\psched_j$ (if it exists) such that $\psched_j \models \varphi$.  All previously considered schedulers are excluded from $\psched$. This ensures that $\psched_j$ is a fresh permissive scheduler; see Section~\ref{sec:computingpermissive} for details. 
\item 
Check all compliant schedulers of $\theta_j$ by reinforcement learning.  This yields scheduler $\sigma^j \in \theta_j$ that minimizes the expected cost of reaching $G$. By Remark~\ref{rem:random_induced} on Page~\pageref{rem:random_induced}, $\sigma_j$ induces a (randomized) scheduler $\sched$ for $\mdp$. The scheduler $\sched$ is safe \wrt $\varphi$ and cost-minimal among all compliant schedulers to $\psched$. During the learning process, the cost function $\rho$ is \emph{refined} with the actual costs for the (newly) explored actions. See Section~\ref{sec:learning} for details.
\item\label{enum:lb} 
Using the refined cost function, a scheduler $\sched_l$ inducing minimal expected cost $c_l$ is computed for the original MDP $\mdp$ (neglecting being safe or not). As this is computed using lower bounds on local costs and potentially using an unsafe scheduler, the expected cost forms a lower bound on cost obtained using full knowledge of the cost function and only safe schedulers.
\item 
After learning the scheduler $\sched$, we check whether $\expRew^{\mdp^{\sched}}(\finally G) \leq \kappa$. Moreover, if the expected cost equals the lower bound computed in Step~\ref{enum:lb}, \ie, $\expRew^{\mdp^{\sched}}(\finally G)=c_l$, the scheduler $\sigma$ is globally optimal (and safe).	
\end{enumerate}

Note that in the worst case, we actually enumerate all possible safe schedulers, \ie the maximal permissive scheduler. However, the iterative nature of the procedure together with the optimizations allows for earlier termination as soon as the optimum is reached or the gap between the lower and upper bounds for the minimal expected cost is sufficiently small.

\begin{theorem}
Safety-constrained reinforcement learning is sound and complete.
\end{theorem}

The method is sound and complete because finally we iterate over all safe permissive schedulers and thereby over all possible safe schedulers.

\subsection{Computing permissive schedulers}\label{sec:computingpermissive}
%	In the first step, for $\mdp$ and $\varphi$ a \emph{safe permissive} scheduler $\psched\in\Psched^\mdp$ is computed. Moreover, all previously computed schedulers are explicitly excluded from the permissive scheduler.
	
In the following, we discuss how to compute a safe deterministic permissive scheduler that induces a safe sub-MDP such as illustrated in Example~\ref{ex:submdp}. Moreover, we indicate how a safe permissive scheduler can be computed in general (for randomized schedulers). Recall that according to our setting we are given an MDP $\MdpInit$ and a safety specification $\varphi=\reachProplT$ for $T\subseteq S$.

The computation will be performed by means of an SMT encoding. This is similar to the mixed linear integer programming (MILP) approach used in~\cite{draeger-et-al-tacas-2014}. The intuition is that a satisfying assignment for the encoding induces a \emph{safe} permissive scheduler according to Definition~\ref{def:safe_scheduler}. We use the following variables.

\begin{description}
	\item [$y_{s,a}\in\mathbb{B} = \{ \true, \false \}$] for each state $s\in S$ and each action $a\in\Act(s)$ is assigned \true iff action $a$ is allowed to be taken in state $s$. These variables form the permissive scheduler.
	\item [$p_s \in\Ireal$] for each state $s\in S$ captures the \emph{maximal} probability to reach the set of target states $T\subseteq S$ under each possible scheduler that is compliant to the permissive scheduler.
\end{description}

The SMT encoding reads as follows.
\begin{align}
											&\quad p_{\sinit}\leq\lambda\label{eq:safe:threshold}\\
	 \forall s\in S.						&\quad \bigvee_{a\in\Act(s)} y_{s,a}\label{eq:safe:scheduler}\\
	 \forall s\in T.						&\quad p_s=1\label{eq:safe:targetprob}\\
	 \forall s\in S.\,\forall a\in\Act(s).	&\quad y_{s,a}\rightarrow p_s\geq\sum_{s'\in S}\pmdp(s,a)(s')\cdot p_{s'}\label{eq:safe:lowerbound}
\end{align}

First, Constraint~\ref{eq:safe:threshold} ensures that the maximal probability at the initial state $\sinit$ achieved by any scheduler that can be constructed according the valuation of the $y_{s,a}$-variables does not exceed the given safety threshold $\lambda$. Due to Constraint~\ref{eq:safe:scheduler}, at least one action $a\in\Act$ is chosen by the permissive scheduler for every state $s\in S$ as at least one $y_{s,a}$-variable needs to be assigned \true. The probability of target states is set to $1$ by Constraint~\ref{eq:safe:targetprob}. Finally, Constraint~\ref{eq:safe:lowerbound} puts (multiple) lower bounds on each state's probability: For all $s\in S$ and $a\in\Act$ with $y_{s,a}=\true$, the probability to reach the target states is computed according to this particular choice and set as a lower bound. Therefore, only combinations of $y_{s,a}$-variables that induce safe schedulers can be assigned \true. 

Now, consider a deterministic scheduler $\sched\in\Sched^\mdp$ which we want to \emph{explicitly exclude} from the computation.
%For every state $s\in S$ there is an action $a\in\Act$ such that $\sched(s)=a$.
It needs to be ensured that for a satisfying assignment at least for one state the corresponding $y_{s,\sigma(s)}$ variable is assigned \false in order to at least make one different decision. This can be achieved by adding the disjunction $\bigvee_{s\in S}\neg y_{s,\sigma(s)}$ to the encoding.

\begin{theorem}
The SMT encoding given by Constraints~\ref{eq:safe:threshold}--\ref{eq:safe:lowerbound} is sound and complete.
\end{theorem}

\paragraph{Proof sketch.} \emph{Soundness} refers to the fact that each satisfying assignment for the encoding induces a safe deterministic permissive scheduler for MDP $\mdp$ and safety specification $\varphi$. This is shown by constructing a permissive scheduler according to an arbitrary assignment of $y_{s,a}$-variables. Applying the other (satisfied) constraints ensures that this scheduler is safe.
\emph{Completeness} means that for each safe deterministic permissive scheduler, a corresponding satisfying assignment of the constraints exists. This is done by assuming a safe deterministic permissive scheduler and constructing a corresponding assignment. Checking all the constraints ensures that this assignment is satisfying.
\smallskip

Using an SMT solver like \tool{Z3}, this encoding does not ensure a certain grade of permissiveness, \ie, that as many $y_{s,a}$-variables as possible are assigned \true. While this is a typical setting for MAX-SMT \cite{maxsmt}, in the current stable version of \tool{Z3} this feature is not available yet. Certain schedulers inducing high probabilities or desired behavior can be  included using the \emph{assumptions} of the SMT solver.

An alternative would be to use an MILP encoding like, \eg, in~\cite{draeger-et-al-tacas-2014,wimmer-et-al-tcs-2014}, and optimize towards a maximal number of available nondeterministic choices. However, in our setting it is crucial to ensure \emph{incrementality} in the sense that if certain changes to the constraints are necessary this does not trigger a complete restart of the solving process.

Finally, there might be safe \emph{randomized schedulers} that induce better optimal costs than all deterministic schedulers~\cite{DBLP:journals/lmcs/EtessamiKVY08,DBLP:conf/csl/BaierDK14}. To compute \emph{randomized permissive schedulers}, the difficulty is that there are arbitrarily (or even infinitely) many probability distributions over actions. A reasonable approach is to bound the number of possible distributions by a fixed number $n$ and introduce for each state $s$, distribution $\distFunc_i$, and action $a$ a real-valued variable $y_{s,\distFunc_i,a}$ for $1\leq i\leq n$. Constraint~\ref{eq:safe:scheduler} is modified such that for all states and actions the $y_{s,\distFunc_i,a}$-variables sum up to one and the probability computation in Constraint~\ref{eq:safe:lowerbound} has to take probabilities over actions into account. Note that the MILP approach from~\cite{draeger-et-al-tacas-2014} cannot be adapted to randomized schedulers as non-linear constraints are involved.

\subsection{Learning}\label{sec:learning}
	 In the learning phase, the main goal of this learning phase is the \emph{exploration} of this MDP, as we thereby learn the cost function. In a more practical setting, we should balance this with \emph{exploitation}, \ie, performing close to optimal---within the bounds of the permissive scheduler---during the learning. 
	 The algorithm we use for the reinforcement learning is \emph{Q-learning}~\cite{qlearning}. To favor the exploration, we initialize the learning with \emph{overly-optimistic} expected rewards. Thereby, we  explore large portions of the MDP while favoring promising regions of the MDP.
	 
Proper balancing of exploration vs. exploitation depends on the exact scenario~\cite{DBLP:journals/jmlr/BrafmanT02}. 
Here, the balance is heavily affected by the construction of permissive schedulers. For instance, if we try to find permissive schedulers which do not exclude the currently best known scheduler, then the exploitation during the learning phase might be higher, while we might severely restrict the exploration.

\section{Experiments}\label{sec:experiments}
We implemented a prototype of the aforementioned synthesis loop in \cpp{} and conducted experiments using case studies motivated by robotic motion planning. Our prototype uses the SMT-based permissive scheduler computation described in Section \ref{sec:computingpermissive} and seeks a \emph{locally} maximal permissive scheduler by successively adding as many actions as possible.

Every MDP considered in the case studies has a set of bad states (that may only be reached with a certain probability) and a set of goal states that the system tries to reach. All case studies feature a relatively large number of nondeterministic choices in each state and a high amount of probabilistic branching to illustrate the applicability of our method to practically relevant models with a high number of schedulers that achieve various performances.

\paragraph{Janitor.}
This benchmark is loosely based on the grid world robot from~\cite{KNP12b}. It features a grid world with a controllable robot. In each step, the robot either changes its direction (while remaining on the same tile) or moves forward in the currently selected direction. Doing so, the robot consumes fuel depending on the surface it currently occupies. The goal is to minimize the fuel consumption for reaching the opposite corner of the grid world while simultaneously avoiding collision with a janitor that moves randomly across the board. 

\paragraph{Following a line fragment.}
We consider a (discretized) variant of a machine that is bound to follow a straight line, \eg a sawmill. In each step, there is a certain probability to deviate from the line depending on the speed the machine currently operates at. That is, higher speeds come at the price of an increased probability to deviate from the center. Given a fixed tolerable distance $d$, the system must avoid to reach states in which the distance from the center exceeds $d$. Also, the required time to complete the task or the required energy are to be minimized, both of which depend on the currently selected speed mode of the system.
	 
\paragraph{Communicating explorer.}
Finally, we use the model of a semi-autonomous explorer as described in \eg~\cite{stuckler2015nimbro}. Moving through a grid-like environment, the system communicates with its controller via two lossy channels for which the probability of a message loss depends on the relay the location of the explorer. The explorer can choose between performing a limited number of attempts to communicate or moving in any direction in each time step. Similarly to the janitor case study, the system tries to reach the opposite corner of the grid while avoiding states in which the explorer moved too far without any (successful) intermediate communication.

For this model, the cost to be optimized is the energy consumption of the electronic circuit, which induces cost for movement, \eg by utilizing sensors, and (significantly higher) cost for utilizing the communcation channels.

\paragraph{Benchmark results.}
Table~1 summarizes the results we obtained using our prototype on a MacBook Pro with an 2.67GHz Intel Core i5 processor and a memory limit of 2GB. As SMT-backend, we used Z3~\cite{dMB08} in version 4.4.0. For several instances of each case study, we list the number of states, transitions, and probabilistic branches. Furthermore, we give the bound $\lambda$ used in the safety property and the optimal performance over all safe schedulers. The following columns provide information about the progress of the synthesis procedure over several selected iterations. The first of these columns ($i$) shows the number of iterations performed thus far, \ie, the number of permissive schedulers on which we applied learning. For iteration $i$, we give the cumulative time $t$ required for the computation of the permissive scheduler as well as the current lower and upper bound on the cost (\wrt the performance measure).

\begin{table}[t]
\centering
\label{tab:benchres}
\scalebox{0.8}{
\begin{tabular}{@{}ll@{\hspace{0.6cm}}rrrrr@{\hspace{0.4cm}}lrrr@{}}
\toprule
Benchmark         &          & states          	& trans. & branch.         & $\lambda$     & \textbf{opt.}    			 & $i$ & $t$ 			&  lower & \textbf{upper} \\ \midrule
\multirow{2}{*}{Janitor} & \multirow{2}{*}{5,5} & \multirow{2}{*}{625} & \multirow{2}{*}{1125}& \multirow{2}{*}{3545} &  \multirow{2}{*}{0.1} & \multirow{2}{*}{\textbf{88.6}} &  1           &  813    		& 84       & \textbf{88.6}  \\ \cmidrule(l){8-11} 
 						& & &                   &                   	&                   &                   			&  2           &  2578    	& 84      & \textbf{88.6}				\\  \cmidrule(l){1-11}

\multirow{6}{*}{FolLine} & \multirow{3}{*}{30,15} & \multirow{3}{*}{455} & \multirow{3}{*}{1265} &\multirow{3}{*}{3693}&  \multirow{3}{*}{0.01} & \multirow{3}{*}{\textbf{716.0}}&  1           &  41    		& 715.4       & \textbf{717.1}  \\ \cmidrule(l){8-11} 
 					& &	&                   &                   	&                   &                   			&  3           &  85    	& 715.62      & \textbf{716.83}				\\  \cmidrule(l){8-11} 
  					& & &                   &                   	&                   &                   				&  13          &  306    	& 715.9          &  \textbf{716.5}				\\  \cmidrule(l){2-11} 
  					& \multirow{3}{*}{40,15} & \multirow{3}{*}{625} 	& \multirow{3}{*}{1775} &\multirow{3}{*}{5223}& \multirow{3}{*}{0.12} & \multirow{3}{*}{\textbf{966.0}} &   1           &  304    		&  964.8         & \textbf{968.2} \\ \cmidrule(l){8-11} 
              & &   &                   &                   	&                   &                   				&  3           &  420    		&  965.4         & \textbf{967.2} \\  \cmidrule(l){8-11} 
              & &    &                   &                   	&                   &                   				&  8           &  738    		&  965.6         & \textbf{966.7} \\  \cmidrule(l){1-11} 

\multirow{7}{*}{ComExp} & \multirow{3}{*}{6,6,6} & \multirow{3}{*}{823} & \multirow{3}{*}{2603} &\multirow{3}{*}{3726}& \multirow{3}{*}{0.08}&  \multirow{3}{*}{\textbf{54.5}} &  1           &  5    	& 0.3       & \textbf{113.3}  \\ \cmidrule(l){8-11} 
 					& & 	&                   &                   	&                   &                   				&  2           &  26    		&  0.3         &  \textbf{74.9}				\\  \cmidrule(l){8-11} 
             & &     &                   &                   	&                   &                   				&  3           &  105    		&  0.3         &  \textbf{57.3}				\\ \cmidrule(l){2-11} 
           & \multirow{4}{*}{8,8,6} & \multirow{4}{*}{1495} 	& \multirow{4}{*}{4859} & \multirow{4}{*}{6953}& \multirow{4}{*}{0.12}& \multirow{4}{*}{\textbf{72.9}}         &  1           &  15    		&  0.42         & \textbf{163.1} \\ \cmidrule(l){8-11} 
            & &      &                   &                   	&                   &                   				&  2           &  80    		&  0.42         & \textbf{122.0} \\ \cmidrule(l){8-11} 
            & &      &                   &                   	&                   &                   				&  3           &  112    		&  0.42         & \textbf{90.1} \\ \cmidrule(l){8-11} 
            & &     &                   &                   	&                   &                   				&  7           &  1319    		&  0.42         & \textbf{78.2}
                  \\\bottomrule\\
\end{tabular}
}
\caption{Benchmark results}
\end{table}

\paragraph{Discussion of the results.}
For the Janitor and FolLine case studies, we observe that the investment of computing a locally maximal permissive scheduler pays off, meaning that we get very tight lower and upper bounds already after the first deployment. This investment comes at the cost of a higher computational effort (per iteration). This could be reduced by more elaborate heuristics which limit our search for (local) maximal permissiveness.

For the communicating explorer, the situation is more difficult. Since a scheduler that does not communicate at all has very low expected costs, a loose lower bound has been obtained. This bound could be severely improved upon by obtaining tighter lower bounds via multi-objective model checking.

\paragraph{Lessons learned.}
The experiments uncovered some intricacies unmentioned by prior work. Computing highly permissive schedulers as in \eg~\cite{draeger-et-al-tacas-2014} often induces small reachable state spaces. This is due to the fact that \emph{all} actions in the unreachable fragments can be selected by the solver.
 It seems that quantifying permissiveness should only consider actually reachable states. This observation is related to the general problem of forcing a solver to ensure reachability of certain states, which would also beneficial for ensuring the reachability of, \eg, goal states. However, any guidance towards this proved to drastically decrease the solver performance.

\section{Conclusion and future work}\label{sec:conclusion}
We presented the---to the best of our knowledge---first approach on iteratively computing safe and optimal strategies in a setting subject to random choices, unknown cost, and safety hazards. Our method was shown to work on practical benchmarks involving a high degree of nondeterminism. Future work will concern improving the scalability by employing multi-objective model checking in order to prove optimality at earlier iterations of the process. Moreover, extensions to stochastic 2-player games for modeling adversarial environment behavior or the investigation of unknown probability distributions seem very interesting. 

\paragraph{Acknowledgements.}	We want to thank Benjamin Lucien Kaminski for the valuable discussion on the worst case size of conflicting sets.

\newpage
\bibliographystyle{splncs}
\bibliography{literature}

\end{document}